\documentclass[aps,prb,twocolumn]{revtex4}
\usepackage{graphicx}
\begin{document}
\title{The inelastic relaxation time due to electron-electron collisions\\ 
in high-mobility two-dimensional systems under microwave radiations}
\author{X.L. Lei and S.Y. Liu}
\affiliation{Department of Physics, Shanghai Jiaotong University,
1954 Huashan Road, Shanghai 200030, China}

\begin{abstract}
In some theoretical analyses of microwave-induced magnetoresistance 
oscillations in high-mobility two-dimensional systems, the "inelastic relaxation 
time" $\tau_{\rm in}$ due to electron-electron scattering is evaluated 
using an equilibrium distribution function $f^0$ in the absence of radiation,  
and it is concluded that $\tau_{\rm in}$ is much larger than $\tau_q$, 
the single-particle relaxation time due to impurity scattering. 
However, under the irradiation of a microwave capable of producing magnetoresistance 
oscillation, the distribution function of the high-mobility electron gas
 deviates remarkably from $f^0$ at low temperatures. 
Estimating $\tau_{\rm in}$ using an approximate nonequilibrium distribution function  
rather than using $f^0$, one will find the system to be in the opposite limit 
$1/\tau_{\rm in}\ll 1/\tau_{q}$ even for $T=0$\,K.
Therefore, models which depend on the assumption $1/\tau_{\rm in}\gg 1/\tau_{q}$ 
may not be justifiable.
\end{abstract}

\maketitle

The thermalization time $\tau_{\rm th}$, i.e. the time needed for system to approach the
thermoequilibrium state, is an important property of an electron system in an nonequilibrium 
state. Though almost all the scattering mechanisms can contribute to $\tau_{\rm th}$, 
in many cases electron-electron scattering is the dominant one for thermalization,
and the "inelastic relaxation time" $\tau_{\rm in}$ due to electron-electron (ee) scattering
in the nonequilibrium state can be approximately considered as $\tau_{\rm th}$.
This $\tau_{\rm in}$ has also served as an important 
parameter in several theoretical models of microwave-induced magnetoresistance 
oscillation in high-mobility two-dimensional systems. These models 
make use of the Boltzmann-type transport equation for the distribution function
$f$ with the ee collision term written in the form
\begin{equation}
\left(\frac{\partial f}{\partial t}\right)_{ee}=\frac{f-f^0}{\tau_{\rm in}},
\end{equation}
where $f^{0}$ is the equilibrium distribution function, i.e. the distribution function
in the absence of 
radiation at lattice temperature $T$. Most of the existing treatments in the 
literature assume, explicitly or implicitly, that the distribution function $f$ 
only slightly deviates from $f^0$, such that $\tau_{\rm in}$ can essentially be 
evaluated using the equilibrium distribution function $f^0$ and get at low 
temperature $T$
\begin{equation}
\frac{1}{\tau_{\rm in}}\approx \frac{T^2}{2\pi E_{\rm F}}\ln\left(\frac{T}{E_{\rm F}}\right)
\end{equation}
with $E_{F}$ the Fermi energy of the two-dimensional (2D) electron system. For GaAs-based 
2D electron system of carrier density $N_{\rm e}=3\times 10^{15}$\,m$^{-2}$, this yields
\[
1/{\tau_{\rm in}}\approx 10\,{\rm mK \,\,\,\, at\,\,\,} T=1\,{\rm K},
\]  
which is of the same order of magnitude as the disorder-limited inverse momentum 
(transport) relaxation time $1/{\tau_{m}}\approx 10\,{\rm mK}$ if the system 
linear mobility is $\mu_0=2000$\,m$^{2}$/Vs, and may be much smaller than the 
inverse single-particle life-time, $1/\tau_{q}$, which can be as large as 0.5\,K. 
The condition $\tau_{\rm in}\gg \tau_{q}$ has been used as a key point 
in many theoretical treatments in the literature.
 
 However, for ultra-clean (high-mobility) 2D electron systems under microwave
 irradiation of strength capable of producing photoresistance oscillation, the 
 nonequilibrium distribution function $f$ remarkably deviates from $f^0$ 
 at low temperatures. The ee collision term in the transport equation generally can 
 not be written in a form of Eq.\,(1). 
 Even if, for an approximate analysis, one still takes Eq.\,(1) in the main equations, 
 $\tau_{\rm in}$ should be calculated using the nonequilibrium distribution function $f$ 
 rather than using $f^0$. Though we may not know the exact nonequilibrium distribution 
 function $f$ under the influence of a strong electric field ${\bf E}_{s}$, it is, in any case, 
 much closer to the shifted function 
 $f^0({\bf k}-m{\bf v}_d)$ than to $f^0({\bf k})$, the equilibrium distribution function
 in the absence of electric field. Here ${\bf k}$ stands for wavevector and 
 ${\bf v}_d$ is the drift velocity under electric field ${\bf E}_{s}$. 
 With this nonequilibrium distribution function we have
\begin{equation} 
\frac{1}{\tau_{\rm in}}\approx \left<\frac{\Delta^2}{4\pi E_{\rm F}}
\left[\ln\left(\frac{\Delta}{E_{\rm F}}\right)-\frac{1}{2}-...\right]\right>,
\end{equation}
 even at $T$=0\,K. Here the excitation energy $\Delta$ can be as large as
\[
\Delta_m=\frac{1}{2}mv_d^2\left(1+2\frac{v_{\rm F}}{v_d}\right).
\]
   For the system of $\mu_0=2000$\,m$^{2}$/Vs subject to an electric field of 
 $E_{s}=0.5$\,V/cm, $v_d \approx 10^5$\,m/s, $\Delta_m\approx 80$\,K 
 and $1/\tau_{\rm in}$
can be as high as 20\,K, with an average around
\[
1/\tau_{\rm in}\approx 10\,\,{\rm K},
\] 
Which is not only much larger than $1/{\tau_{m}}\approx 10\,{\rm mK}$ but also 
much larger than $1/{\tau_{q}}\approx 0.5\,{\rm K}$.

Therefore, models which depend heavily on the assumption $1/\tau_{\rm in}\gg 1/\tau_{q}$ 
may not be justifiable, even without considering other mechanisms for system thermalization.

\end{document}